\documentclass[pre,twocolumn,twoside,showpacs,byrevtex,superscriptaddress,floatfix]{revtex4}

\usepackage{currfile}
\usepackage{graphicx,epsfig,verbatim,enumerate}
\usepackage{amssymb,amsmath,mathdots}
\usepackage{ifthen}

\usepackage{tikz-cd}
\usetikzlibrary{arrows}
\tikzset{
  commutative diagrams/.cd,
  arrow style=tikz,
  diagrams={>=space}}

\newboolean{twocolswitch}

\newcommand{\sindex}[1]{}
\newcommand{\nindex}[1]{}

\newcommand{\www}[1]{\url{#1}}

\usepackage{lettrine}

\usepackage{color}

\setboolean{twocolswitch}{true}
\newcommand{\revtexonly}[1]{#1}

\begin{document}

\title{
  
Characterizing the Google Books corpus:\\
Strong limits to inferences of socio-cultural and linguistic evolution

}

\author{
\firstname{Eitan Adam}
\surname{Pechenick}
}

\email{eitan.pechenick@uvm.edu}

\affiliation{
  Computational Story Lab,
  Vermont Complex Systems Center,
  Vermont Advanced Computing Core,
  \& the Department of Mathematics and Statistics,
  University of Vermont,
  Burlington,
  VT, 05401
}

\author{
\firstname{Christopher M.}
\surname{Danforth}
}
\email{chris.danforth@uvm.edu}

\affiliation{
  Computational Story Lab,
  Vermont Complex Systems Center,
  Vermont Advanced Computing Core,
  \& the Department of Mathematics and Statistics,
  University of Vermont,
  Burlington,
  VT, 05401
}

\author{
\firstname{Peter Sheridan}
\surname{Dodds}
}
\email{peter.dodds@uvm.edu}

\affiliation{
  Computational Story Lab,
  Vermont Complex Systems Center,
  Vermont Advanced Computing Core,
  \& the Department of Mathematics and Statistics,
  University of Vermont,
  Burlington,
  VT, 05401
}

\date{\today}

\begin{abstract}
  
It is tempting to treat frequency trends from the Google Books data sets
as indicators of the ``true'' popularity of various words and
phrases. Doing so allows us to draw quantitatively strong conclusions about the
evolution of cultural perception of a given topic, such as time or
gender.
However, the Google Books corpus suffers from
a number of limitations which make it an obscure mask of cultural popularity.
A primary issue is that the corpus is in effect a library,
containing one of each book.
A single, prolific author is thereby able to noticeably
insert new phrases into the Google Books lexicon, whether the author
is widely read or not.
With this understood, the Google Books corpus remains an important
data set to be considered more lexicon-like than text-like.
Here, we show that a distinct problematic feature arises from the inclusion of scientific
texts, 
which have become an increasingly substantive portion of the corpus
throughout the 1900s.
The result is a surge of phrases typical to
academic articles but less common in general, such as references to
time in the form of citations.
We use information theoretic methods to highlight these dynamics by
examining and comparing major contributions via 
a divergence measure of English data sets between decades in the period
1800--2000.
We find that only the English Fiction data set from the
second version of the corpus is not heavily affected by professional
texts.
Overall, our findings call into question the vast majority of existing 
claims drawn from the Google Books corpus, 
and point to the need to fully characterize the dynamics of 
the corpus before
using these data sets to draw broad conclusions about cultural and
linguistic evolution.

\end{abstract}

\maketitle

\section{Introduction}
\label{sec:googlebooks.intro}

\lettrine[nindent=0em,lines=2]{T}{}he Google Books data set is
captivating both for its availability and its incredible size. The
first version of the data set, published in 2009, incorporates over 5
million books~\cite{michel2011quantitative}. 
These are, in turn, a
subset selected for quality of optical character recognition and
metadata---e.g., dates of publication---from 15 million digitized
books, largely provided by university libraries. These 5
million books contain over half a trillion words, 361 billion of which
are in English. Along with separate data sets for American English,
British English, and English Fiction; the first version also includes
Spanish, French, German, Russian, Chinese, and Hebrew data sets. The
second version, published in 2012~\cite{lin2012syntactic}, contains 8
million books with half a trillion words in English alone, and also
includes books in Italian. The contents of the sampled books are split
into case-sensitive $n$-grams 
which are typically blocks of text separated into
$n$ = 1, \ldots, 5
pieces by whitespace---e.g., ``I'' is a 1-gram, and ``I~am'' is a
2-gram

A central if subtle and deceptive 
feature of the Google Books corpus, and for others composed
in a similar fashion, is that the corpus is a reflection of 
a library in which only one of each book is available.
Ideally, we would be able to apply different popularity filters
to the corpus.  
For example, we could ask to have $n$-gram frequencies
adjusted according to book sales in the UK, library usage data
in the US, or how often each page in each book is read on Amazon's
Kindle service (all over defined periods of time).
Evidently, incorporating popularity in any useful fashion
would be an extremely difficult undertaking on the part of Google.

We are left with the fact that 
the Google Books library has ultimately been furnished by the efforts and choices 
of authors, editors, and publishing houses, who collectively aim
to anticipate or dictate what people will read.  
This adds a further distancing from ``true culture'' as
the ability to predict cultural success is often rendered fundamentally 
impossible due to social influence processes~\cite{salganik2006a}---we
have one seed for each tree but no view of the real forest that will emerge.

We therefore observe that the Google Books corpus encodes only a
small-scale kind of popularity: how often $n$-grams appear in a library 
with all books given (in principle) equal importance and tied to their year of
publication (new editions and reprints allow some books to appear more
than once).  
The corpus is thus more akin to a lexicon for a collection of
texts, rather than the collection itself.
But problematically, because Google Books $n$-grams do have frequency of usage
associated with them based on this small-scale popularity, 
the data set readily conveys an illusion of large-scale cultural popularity.
An $n$-gram which declines in usage frequency over time may in fact
become more often read by a particular demographic focused on a
specific genre of books.
For example, ``Frodo'' first appears in the second Google Books English Fiction
corpus in the mid 1950s and declines thereafter in popularity with a few
resurgent spikes~\cite{frodo-google-ngrams2015a}.

While this limitation to small-scale popularity tempers the kinds of conclusions we can draw,
the evolution of $n$-grams within the Google Books corpus---their relative abundance, their growth
and decay---still gives us a valuable lens into how language
use and culture has changed over time.
Our contribution here will be to show:
\begin{enumerate}
\item 
  A principled approach for exploring word and phrase evolution;
\item 
  How the Google Books corpus is challenged 
  in other respects orthogonal to its library-like nature, particularly by the inclusion
  of scientific and medical journals; 
  and
\item 
  How future analyses of the Google Books corpus should be considered.
\end{enumerate}

\begin{figure}[tbp!]
  \begin{center}
    \includegraphics[width=0.48\textwidth]{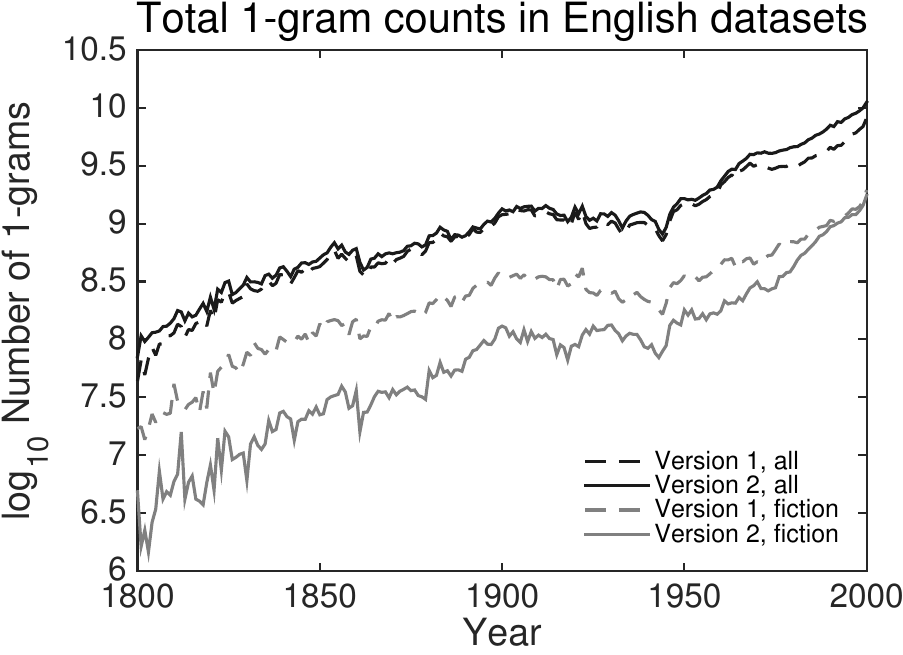}
  \end{center}
  \caption{ 
    The logarithms of the total 1-gram counts for
    the Google Books English data sets (dark gray) and English Fiction data sets (light gray).
    The dashed and solid curves
    denote the 2009 and 2012 versions of the data sets. 
    In all four examples, an exponential increase in volume is apparent over
    time with notable exceptions during wartime when the total volume
    decreases, clearest during the American Civil War and
    both World Wars.  
    While the total volume for English
    increases between versions, the volume for English fiction decreases
    drastically, suggesting a more rigorous filtering process.
  }
  \label{fig:volume}
\end{figure}

For ease of comparison with related work, we focus primarily on
1-grams from selected English data sets between the years 1800 and
2000. In this work, we will use the terms ``word'' and ``1-gram''
interchangeably for the sake of convenience. The total volume of
(non-unique) English 1-grams grows exponentially between these years,
as shown in Fig.~\ref{fig:volume}, except during major conflicts---e.g.,
the American Civil War and both World Wars---when the total volume
dips substantially.  We also observe a slight increase in volume
between the first and second version of the unfiltered English data
set. Between the two English Fiction data sets, however, the total
volume actually decreases considerably, which indicates
insufficient filtering was used in producing the first version, 
and
immediately suggests the initial English Fiction data set may not be
appropriate for any kind of analysis.

The simplest possible analysis involving any Google Books data set is
to track the relative frequencies of a specific set of words or
phrases. Examples of such analyses involve words or phrases
surrounding individuality~\cite{twenge2012increases},
gender~\cite{twenge2012male},
urbanization~\cite{greenfield2013changing}, and
time~\cite{michel2011quantitative,petersen2012statistical}, all of
which are of profound interest. However, the strength of all
conclusions drawn from these must take into account both the number of
words and phrases in question (anywhere from
two~\cite{greenfield2013changing} to twenty~\cite{twenge2012increases}
or more at a time) and the sampling methods used to build the Google
Books corpus.

Many researchers have carried out broad analyses of the Google Books
corpus, examining properties and dynamics of entire languages. These
include analyses of Zipf's and Heaps' laws as applied to the
corpus~\cite{gerlach2013stochastic}, the rates of verb
regularization~\cite{michel2011quantitative}, rates of word
introduction and obsolescence and durations of cultural
memory~\cite{petersen2012statistical}, as well as an observed decrease
in the need for new words in several
languages~\cite{petersen2012languages}. However, these studies also
appear to take for granted that the data sets sample in a consistent
manner from works spanning the last two centuries.

Analysis of the emotional content of books suggests a lag of roughly a
decade between exogenous events and their effects in
literature~\cite{bentley2014books},
complicating the use of the Google Books data sets directly
as snapshots of cultural identity. 

\begin{figure}[tbp!]
  \begin{center}
    \includegraphics[width=0.48\textwidth]{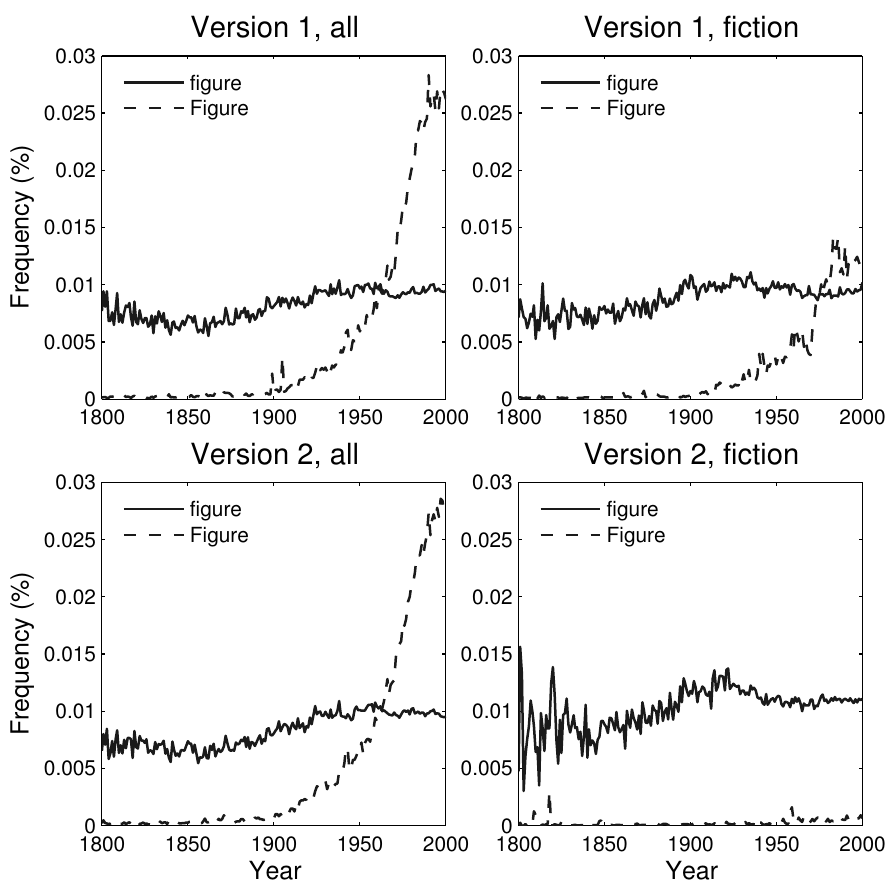}
  \end{center}
  \caption{ 
    Relative frequencies of ``Figure'' vs ``figure'' in both
    versions of the Google Books corpus for both English (all) and
    English Fiction. In the English data sets, the capitalized term
    rapidly surpasses the uncapitalized term in the 1960s. For the first
    English Fiction data set, this effect is delayed until the
    1970s.
    As shown later, only the second version of the English Fiction data set
    demonstrates a filtering 
    of scientific terminology.
    These trends strongly suggest an increase starting around 1900
    in the sampling of scientific texts in both English data sets and
    the first English Fiction data set.
  }
  \label{fig:fFig}
\end{figure}

As we will demonstrate, an assumption of unbiased sampling of books is
not reasonable during the last century and especially during recent
decades, which is of particular importance to all analyses concerned
with recent social change. Since parsing in the data sets is
case-sensitive, we can give a suggestive illustration of this
observation in Fig.~\ref{fig:fFig}, which displays the relative
(normalized) frequencies of ``figure'' versus ``Figure'' in both versions of the
corpus and for both English and English Fiction. In both versions of
the English data set, the capitalized version, ``Figure,'' surpasses
its lowercase counterpart during the 1960s. Since the majority of
books in the corpus originated in university
libraries~\cite{michel2011quantitative}, a major effect of scientific
texts on the dynamics of the data set is quite plausible. This trend
is also apparent---albeit delayed---in the first version of the
English Fiction data set, which again suggests insufficient filtering
during the compilation process for that version.

Because of Google Books library-like nature, authors are not
represented equally or by any measure of popularity
in any given data set but are instead roughly
by their own prolificacy.
This leaves room for individual authors to
have noteworthy effects on the dynamics of the data sets, as we will
demonstrate in Section~\ref{sec:discussion}.

Lastly, due to copyright laws, the public data sets do not include
metadata (see supporting online
material~\cite{michel2011quantitative}), and the data are truncated to
avoid inference of authorship, which severely limits any analysis of
censorship~\cite{michel2011quantitative,koplenig2015a} in the
corpus. Under these conditions, we will show that much caution must be
used when employing these data sets---with a possible exception of the
second version of English Fiction---to draw cultural conclusions from
the frequencies of words or phrases in the
corpus.

We structure the remainder of the paper as follows. In
Sec.~\ref{sec:methods}, we describe how to use Jensen-Shannon
divergence to highlight the dynamics over time of both versions of the
English and English Fiction data sets, paying particular attention 
to key contributing words. 
In Sec.~\ref{sec:discussion}, we display and discuss examples of
these highlights, exploring the extent of the scientific literature
bias and issues with individual authors; we also
provide a detailed inspection of some example decade--decade comparisons.
We offer concluding remarks in Section~\ref{sec:conc}.

\section{ Methods} 
\label{sec:methods}

\subsection{Statistical divergence between years}

We examine the dynamics of the Google Books corpus by calculating the statistical
divergence between the distributions of 1-grams in two given years. A
commonly used measure of statistical divergence is Kullback-Leibler
(KL) divergence~\cite{kullback1951information}, based on which we use
a bounded, symmetric measure. Given a language with $N$ unique words
and 1-gram distributions $P$ in the first year and $Q$ in second, the
KL divergence between $P$ and $Q$ can be expressed as
\begin{equation}
  D_{KL}(P\,||\,Q)  
  = 
  \sum_{i=1}^N 
  p_i
  \log_{2}\frac{p_i}{q_i},
\end{equation}
where $p_i$ is the probability of observing the $i^\text{th}$ 1-gram
random chosen from the 1-gram distribution for first year, and $q_i$ is
the probability of observing the same word in the second year.
The units of KL divergence is bits,
and may be interpreted as the average
number of bits wasted if a text from the first year is encoded
efficiently, but according to the distribution from the latter,
incorrect year. 
To demonstrate this, we may rewrite the previous equation as
\begin{equation}
D_{KL}(P\,||\,Q)  
=
-\sum_ {i=1}^{N}
p_i\log_{2} q_i
-
H(P),
\end{equation}
where $H(P)=-\sum_ip_i\log_{2} p_i$ is the Shannon
entropy~\cite{shannon2001mathematical}, also the average number of
bits required per word in an efficient encoding for the original
distribution; and the remaining term is the average number of bits
required per word in an efficient, but mistaken, encoding of a given
text. However, if a single (say, the $j^\text{th}$) 1-gram in the
language exists in the first year, but not in the second, then
$q_j=0$, and the divergence diverges. Since this scenario is not
extraordinary for the data sets in question, we instead use
Jensen-Shannon divergence (JSD)~\cite{lin1991divergence} given by
\begin{equation}
  D_{\textrm{JS}}(P\,||\,Q) 
  =
  \frac12\big(D_{KL}(P||M)+D_{KL}(Q||M)\big),
\end{equation} 
where $M=\frac12(P+Q)$ is a mixed distribution of the two years. This
measure of divergence is bounded between 0 when the distributions are
the same and 1 bit in the extreme case when there is no overlap
between the 1-grams in the two distributions. 
If we begin with a uniform distribution of $N$ species and replace $k$
of those species with $k$ entirely new ones, the JSD between the
original and new distribution is $k/N$, the proportion of species
replaced. The JSD is also symmetric, which is an added
convenience. The JSD may be expressed as
\begin{equation} \label{eq:jsd}
D_{\textrm{JS}}(P\,||\,Q) = H(M) - \frac12\big(H(P)+H(Q)\big),
\end{equation}
from which it is apparent that a similar waste analogy holds as with KL divergence, with the mixed distribution taking the place of the approximation regardless of the year a text was written.

\subsection{Key contributions of individual words}

\begin{figure}[tbp!]
  \begin{center}
    \includegraphics[width=0.35\textwidth]{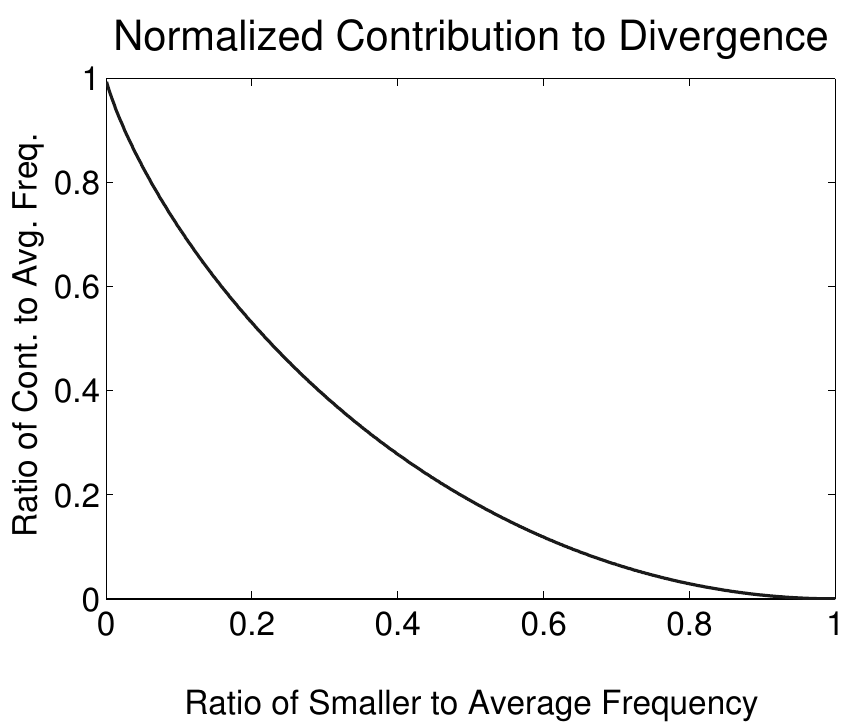}
  \end{center}
  \caption{
    For the ratio $r$ between the smaller relative probability of an
    element and the average, $C(r)$ is the proportion of the average
    contributed to the Jensen-Shannon divergence (see
    Eqs.~\ref{eq:contribution-messy}~and~\ref{eq:contribution}). In
    particular, if $r=1$ (no change), then the contribution is zero; if
    $r=0$, the contribution is half its probability in the distribution in
    which it occurs with nonzero probability.
  }
  \label{fig:contribution_curve}
\end{figure}

The form for Jensen-Shannon divergence given in Eq.~\ref{eq:jsd} can
be broken down into contributions from individual words, where the
contribution from the $i^\text{th}$ word to the divergence between two
years is given by
\begin{equation}
D_{\textrm{JS},i}(P\,||\,Q) = -m_i\log_{2} m_i + \frac12\big(p_i\log_{2} p_i+q_i\log_{2} q_i\big).\end{equation}
Some rearrangement gives
\begin{equation}
D_{\textrm{JS},i}(P\,||\,Q) = m_i \cdot \frac12\big(r_i\log_{2} r_i +
(2-r_i)\log_{2}(2-r_i)\big),
\label{eq:contribution-messy}
\end{equation}
where $r_i=p_i/m_i$, so that contribution from an individual word is
proportional to the average probability of the word, and the proportion
depends on the ratio between the smaller probability (without loss of
generality) and the average. Namely, we may reframe the equation above
as
\begin{equation}
  D_{\textrm{JS},i}(P\,||\,Q) = m_iC(r_i).
  \label{eq:contribution}
\end{equation}
Words with larger average probability yield greater contributions as do
those with smaller ratios, $r$, between the smaller and average
probability.
So while a common 1-gram---such as ``the,'' ``if,'' or a
period---changing subtly can have a large effect on the divergence, so
can an uncommon (or entirely new) word given a sufficient shift from
one year to the next. The size of the contribution relative to the
average probability is displayed in Fig.~\ref{fig:contribution_curve} for
ratios ranging from 0 to 1. $C(r_i)$ is symmetric about $r_i=1$ (i.e.,
no change), so no novel behavior is lost by omitting the case where
$r_i>1$ (i.e., when $p_i$ is the larger probability). The maximum
possible contribution (in bits) is precisely the average probability of
the word in question, which occurs if and only if the smaller
probability is 0. No contribution is made if and only if the probability
remains unchanged.

We coarse-grain the data at the level of decades---e.g., between
1800-to-1809 and 1990-to-1999---by averaging the relative normalized frequency of
each unique word in a given decade over all years in that
decade. (Each year is weighted equally.) This allows convenient
calculation and sorting of contributions to divergence of individual
1-grams between any two time periods.

\section{Results and Discussion} 
\label{sec:discussion}

\subsection{Broad view of language evolution within Google Books}
\label{subsec:broadview}

\begin{figure}[tbp!]
  \begin{center}
    \includegraphics[width=0.48\textwidth]{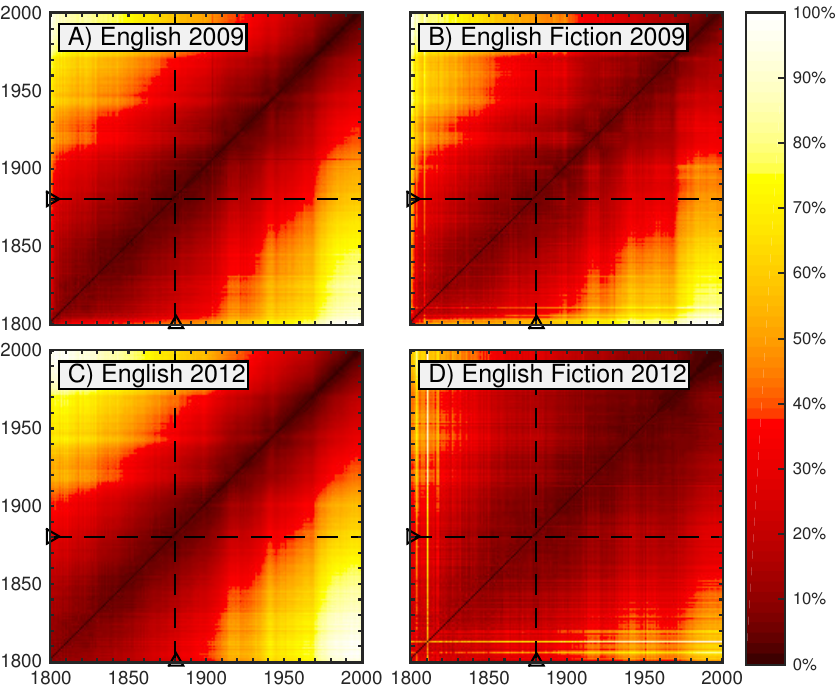}
  \end{center}
  \caption{
    Heatmaps showing the JSD between every pair of years between
    1800 and 2000, contributed by words appearing above a normalized frequency
    threshold of $10^{-5}$. The dashed lines highlight the divergences
    to and from the year 1880, which are featured in
    Fig.~\ref{fig:1g_divcuts}. The off-diagonal elements represent
    divergences between consecutive years, as in
    Fig.~\ref{fig:1g_divcuts2}. The color represents the percentage of the
    maximum divergence observed in the given time range for each data
    set. The divergence between a year and itself is zero. For any given
    year, the divergence increases with the distance (number of years)
    from the diagonal---sharply at first, then gradually. Interesting
    features of the maps are the presence of two cross-hairs in the
    first half of the 20th century, which strongly suggests a wartime
    shift in the language, as well as an asymmetry that suggests a
    particularly high divergence between the first half century and the
    last quarter century observed.
  }
  \label{fig:1g_heatmaps}
\end{figure}

Fig.~\ref{fig:1g_heatmaps} shows the JSD between the 1-gram distributions
for every pair of years between 1800 and 2000 contributed by 1-grams
present above a threshold normalized frequency of $10^{-5}$ for both
versions of the English and English Fiction data sets (i.e., words
that appear with normalized frequency at least 1 in $10^5$).

\begin{figure}[tbp!]
  \begin{center}
    \includegraphics[width=0.48\textwidth]{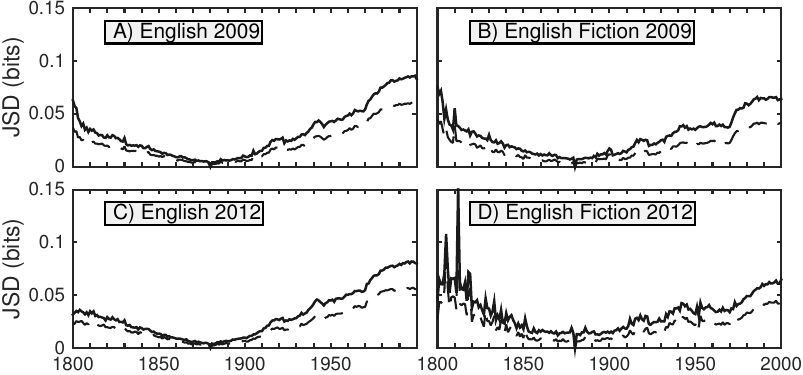}
  \end{center}
  \caption{ 
    JSD between 1880 and each displayed year for given data
    set, corresponding to dashed lines from
    Fig.~\ref{fig:1g_heatmaps}. Contributions are counted for all words
    appearing above a $10^{-5}$ threshold in a given year; for the
    dashed curves, the threshold is $10^{-4}$. Typical behavior in
    each case consists of a relatively large jump between one year and
    the next with a more gradual rise afterward (in both
    directions). Exceptions include wartime, particularly the two
    World Wars, during which the divergence is greater than usual;
    however, after the conclusion of these periods, the cumulative
    divergence settles back to the previous trend. Initial spikiness
    in (D) is likely due to low volume.
  }
  \label{fig:1g_divcuts}
\end{figure}

\begin{figure}[tbp!]
  \begin{center}
    \includegraphics[width=0.48\textwidth]{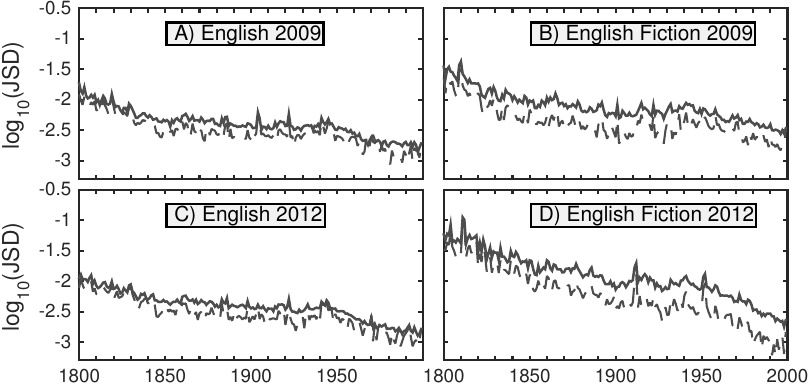}
  \end{center}
  \caption{
    Consecutive year (between each year and the following year)
    base-10 logarithms of JSD, corresponding to off-diagonals in
    Fig.~\ref{fig:1g_heatmaps}. For the solid curves, contributions are
    counted for all words appearing above a $10^{-5}$ threshold in a
    given year; for the dashed curves, the threshold is
    $10^{-4}$. Divergences between consecutive years typically decline
    through the mid-19th century, remain relatively steady until the
    mid-20th century, then continue to decline gradually over time.
  }
  \label{fig:1g_divcuts2}
\end{figure}

A major qualitative aspect apparent from the heatmaps is a gradual
increase in divergence with differences in time---the lexicon
underlying Google steadily evolves---though this is strongly
curtailed for the second English Fiction corpus.
We see the heatmaps are ``pinched'' toward the
diagonal in the vicinities of the two world wars. 
Also visible is an
asymmetry that suggests a particularly high divergence between the
first half century and the last quarter century observed. 
We examine
these effects more closely in
Figs.~\ref{fig:1g_divcuts} and~\ref{fig:1g_divcuts2} by taking 
two slices of the heatmaps.
We specifically consider the divergences of each
year compared with 1880 (dashed lines),
and the divergences between consecutive years (off-diagonal).
To verify qualitative consistency, we also include
analogous contribution curves using the more restrictive threshold
of $10^{-4}$.

While the initial divergence between any two consecutive years is
noticeable, the divergence increases (for the most part) steadily with
the time difference. The cross-hairs from the heatmap resolve into
war-time bumps in divergence, which quickly settle in peacetime. The
larger boost to the divergence in recent decades, however, is more
persistent suggesting a more fundamental change in the data set, which
we will examine in more depth later in this section. Divergences
between consecutive years typically decline through the mid-19th
century. Divergences then remain relatively steady until the mid-20th
century, then continue to decline gradually over time, which may be
consistent with previous findings of decreased rates of word
introduction and increased rates of word obsolescence in many Google
Books data sets over time~\cite{petersen2012statistical} and a slowing
down of linguistic evolution over time as the vocabulary of a language
expands~\cite{petersen2012languages}. The initial spikes in divergence
in the second version of the fiction data set are likely due to the
lower initial volume observed in Fig.~\ref{fig:volume}.

\begin{figure}[tbp!]
  \includegraphics[width=0.48\textwidth]{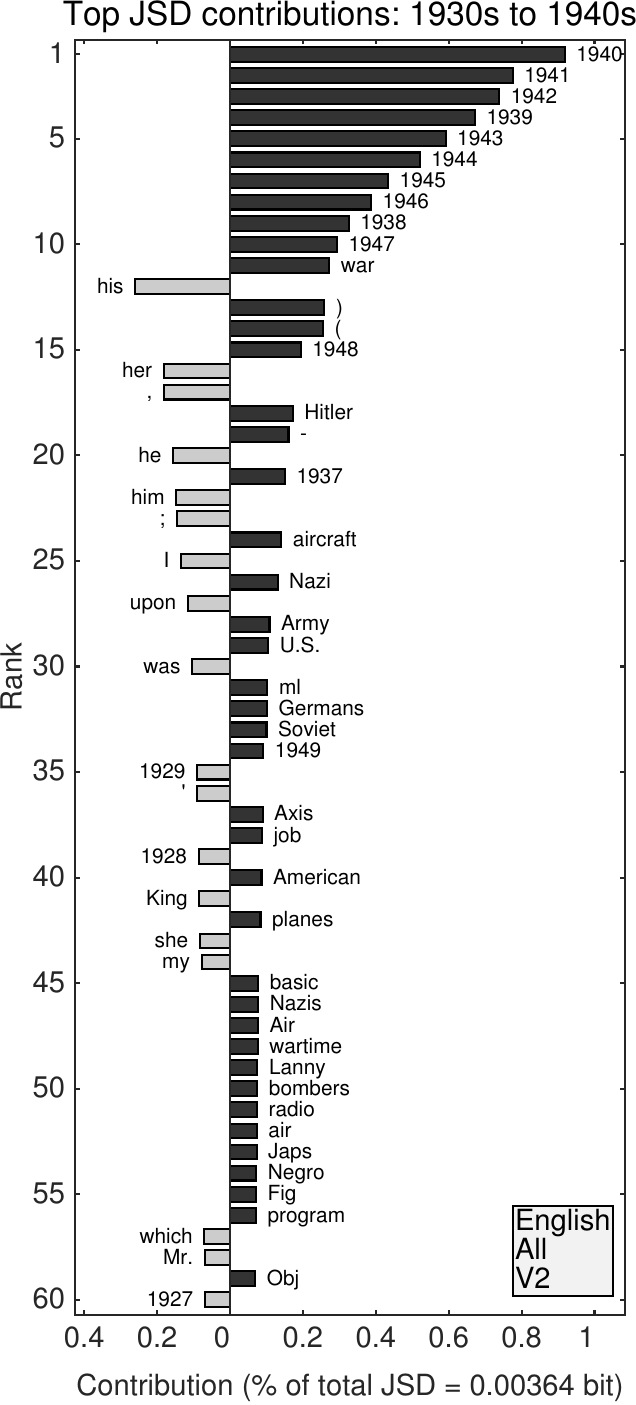}
  \caption{
    (English, all; Version 2.) Top 60 individual contributions of
    1-grams to the JSD between the 1930s and the 1940s. Each
    contribution is given as a percentage of the total JSD (see
    horizontal axis label) between the two given decades. All
    contributions are positive; bars to the left of center represent
    words that were more common in the earlier decade, whereas bars to
    the right represent words that became more common in the later
    decade.
  }
  \label{fig:barsalolnew30}
\end{figure}

\begin{figure}[tbp!]
  \includegraphics[width=0.48\textwidth]{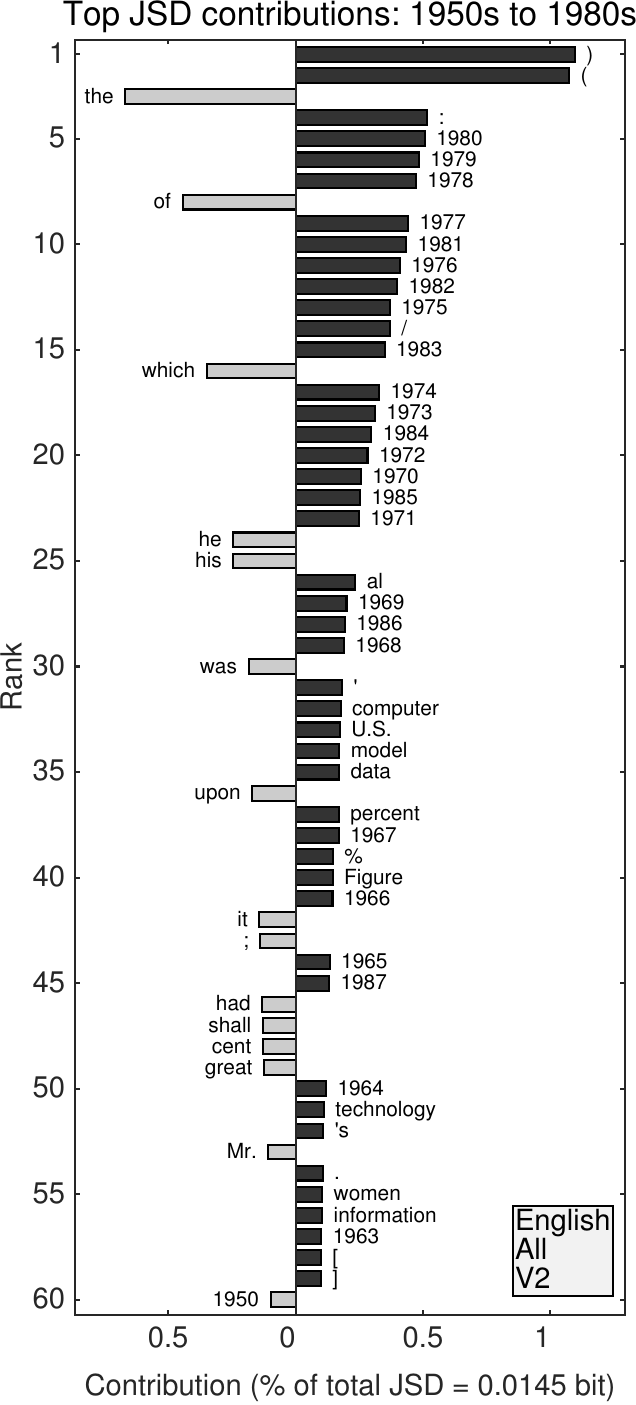}
  \caption{ 
    (English, all; Version 2.) Top 60 individual contributions
    of 1-grams to the JSD between the 1950s and the 1980s. Each
    contribution is given as a percentage of the total JSD (see
    horizontal axis label) between the two given decades. All
    contributions are positive; bars to the left of center represent
    words that were more common in the earlier decade, whereas bars to
    the right represent words that became more common in the later
    decade.
  }
  \label{fig:barsallnew50}
\end{figure}

\begin{figure}[tbp!]
  \includegraphics[width=0.48\textwidth]{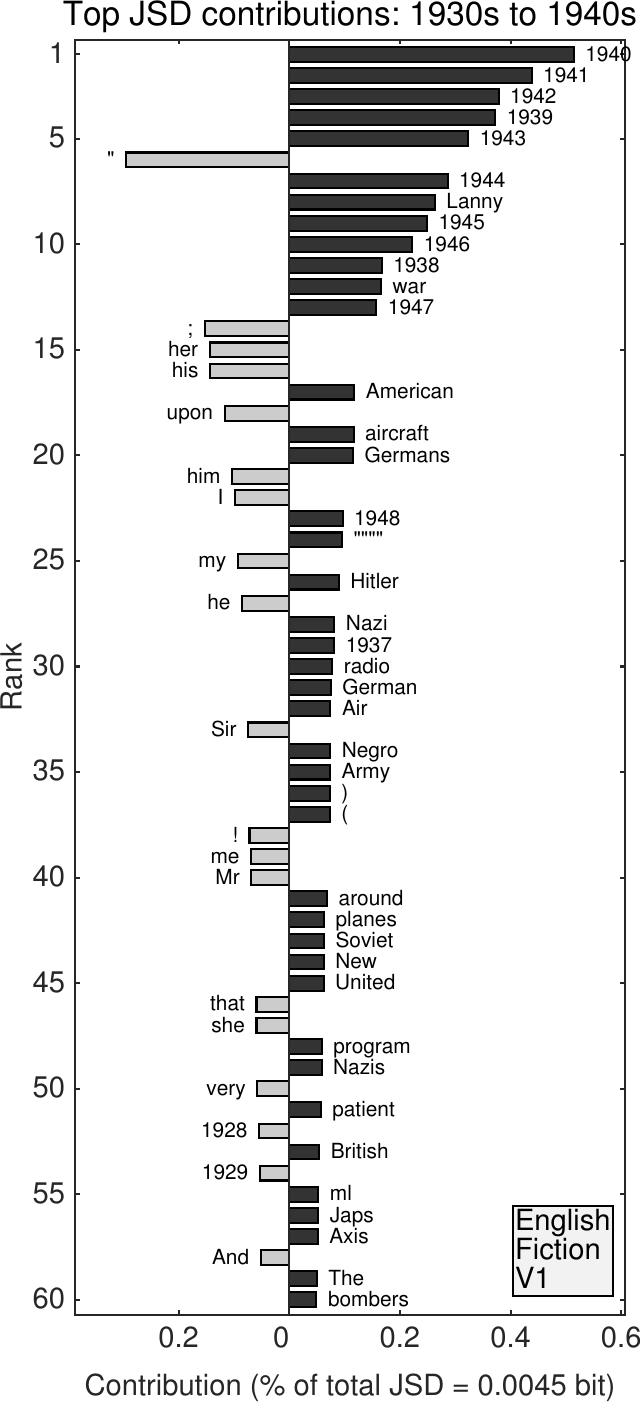}
  \caption{
    (English Fiction, Version 1.) Top 60 individual contributions of
    1-grams to the JSD between the 1930s and the 1940s. Each
    contribution is given as a percentage of the total JSD (see
    horizontal axis label) between the two given decades. All
    contributions are positive; bars to the left of center represent
    words that were more common in the earlier decade, whereas bars to
    the right represent words that became more common in the later
    decade.}
  \label{fig:barsficold30}
\end{figure}

\begin{figure}[tbp!]
  \includegraphics[width=0.48\textwidth]{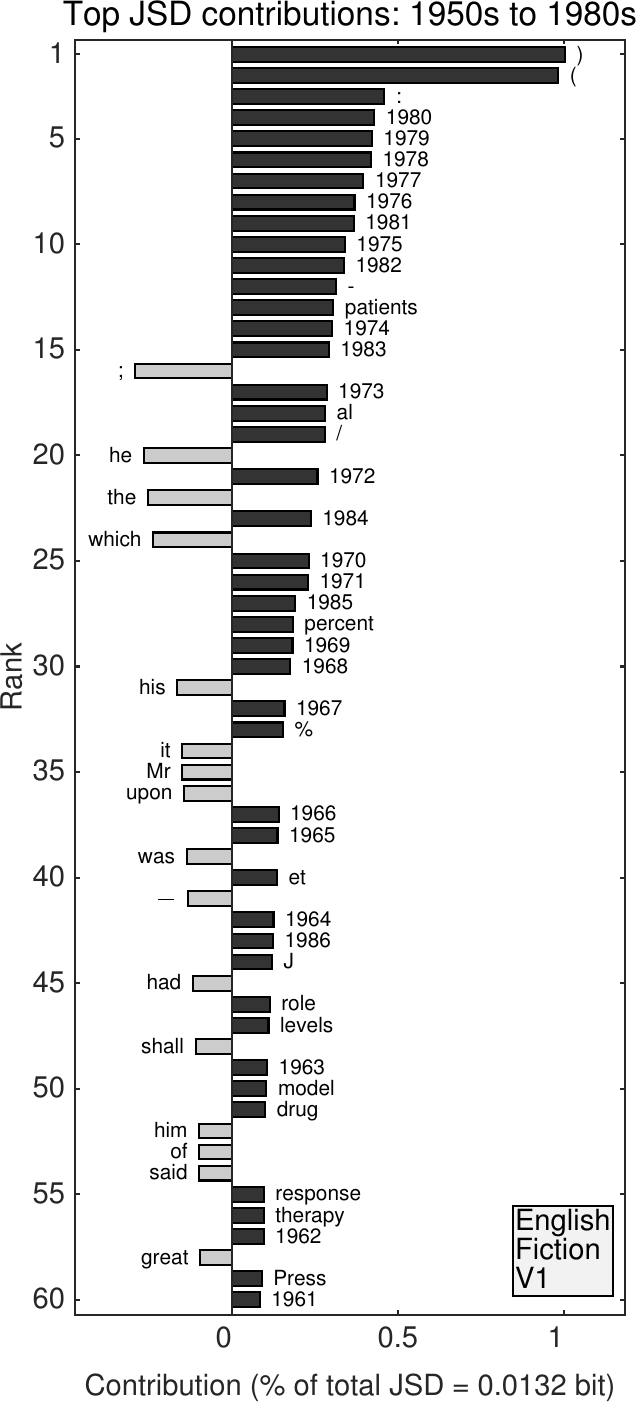}
  \caption{
    (English Fiction, Version 1.) Top 60 individual contributions of
    1-grams to the JSD between the 1950s and the 1980s. Each
    contribution is given as a percentage of the total JSD (see
    horizontal axis label) between the two given decades. All
    contributions are positive; bars to the left of center represent
    words that were more common in the earlier decade, whereas bars to
    the right represent words that became more common in the later
    decade.
  }
  \label{fig:barsficold50}
\end{figure}

\begin{figure}[tbp!]
  \includegraphics[width=0.48\textwidth]{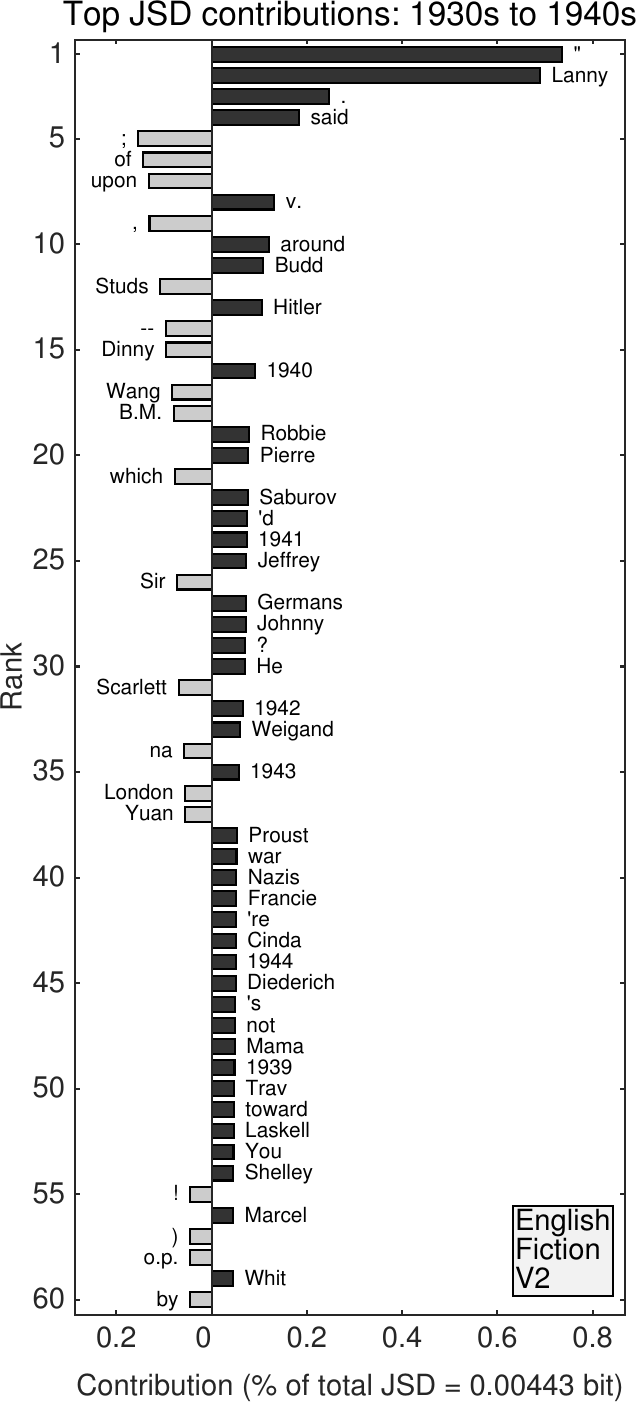}
  \caption{
    (English Fiction, Version 2.) Top 60 individual contributions
    of 1-grams to the JSD between the 1930s and the 1940s. Each
    contribution is given as a percentage of the total JSD (see
    horizontal axis label) between the two given decades. All
    contributions are positive; bars to the left of center represent
    words that were more common in the earlier decade, whereas bars to
    the right represent words that became more common in the later
    decade.}
  \label{fig:barsficnew30}
\end{figure}

\begin{figure}[tbp!]
  \includegraphics[width=0.48\textwidth]{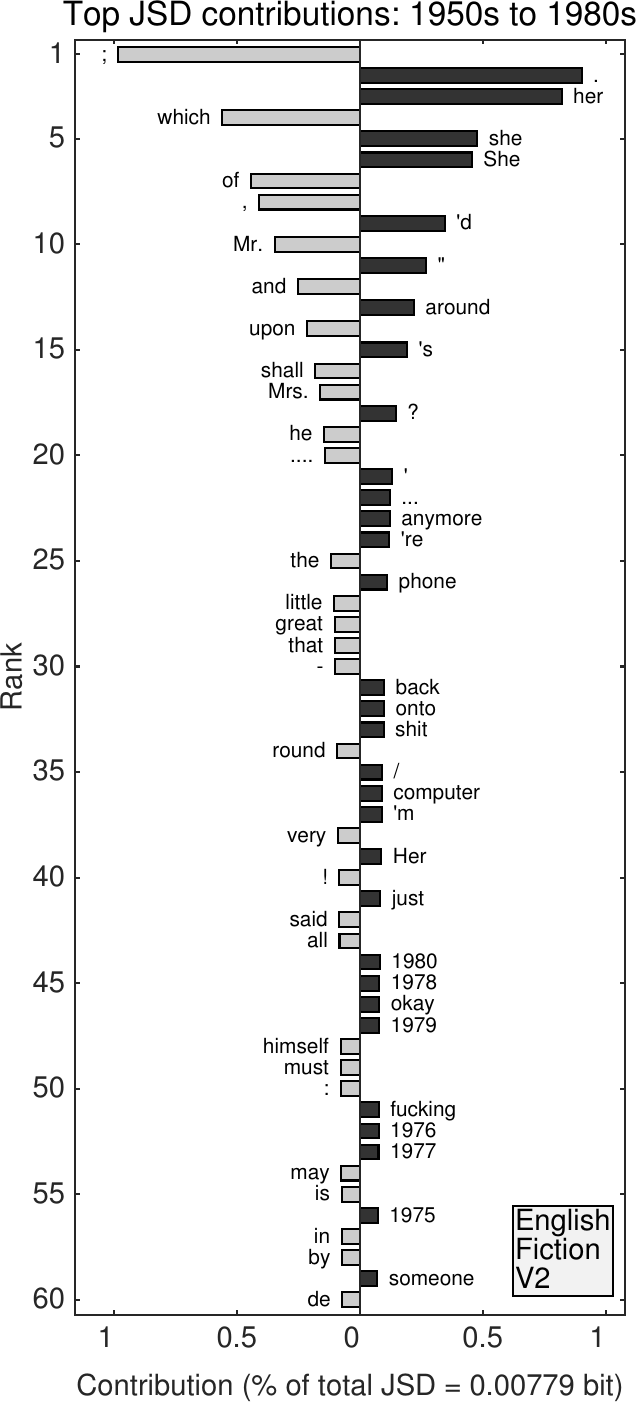}
  \caption{
    (English Fiction, Version 2.) Top 60 individual contributions
    of 1-grams to the JSD between the 1950s and the 1980s. Each
    contribution is given as a percentage of the total JSD (see
    horizontal axis label) between the two given decades (see
    title). All contributions are positive; bars to the left of center
    represent words that were more common in the earlier decade, whereas
    bars to the right represent words that became more common in the
    later decade.
  }
  \label{fig:barsficnew50}
\end{figure}

\subsection{Decade-decade comparisons using JSD word shifts}
\label{subsec:wordshifts}

\subsubsection{General observations}
\label{subsubsec:wordshifts-general}

We present ``word shifts''~\cite{dodds2011e}
for a few examples of inter-decade divergences 
in Figs.~\ref{fig:barsalolnew30}--\ref{fig:barsficnew50},
specifically comparing the 1940s to the 1930s and the 1980s to the
1950s
for the second unfiltered English data set
(Figs.~\ref{fig:barsalolnew30}--\ref{fig:barsallnew50})
and both English Fiction data sets
(Figs.~\ref{fig:barsficold30}--\ref{fig:barsficnew50}).
We provide a full set of such comparisons in the supplementary S1--S4 Files.
For each of the four data sets,
the largest contributions to all divergences generally appear to be
from increased relative frequencies of use of words between
decades.  
For the unfiltered data sets, these are in turn heavily
influenced by increased mention of years, 
which is less pronounced for English Fiction.

The 1940s literature, unsurprisingly, features
more references to Hitler and war than the 1930s, along with other
World War II-related military and political terms.
This is seen regardless of the specific data set used and is fairly
encouraging. Curiously, regardless of the specific data set, a
noticeable contribution is given by an increase in relative use of the
words ``Lanny'' and ``Budd,'' in reference to one character (Lanny
Budd) frequently written about by Upton Sinclair during that
decade. In the fiction data sets, this character dominates
the charts.

\subsubsection{Second unfiltered English data set: 1930s versus 1940s}
\label{subsubsec:wordshifts-unfeng2-1930s-1940s}

A comparison of the 1930s and 1940s for the 
second version of the unfiltered English data set 
(Fig.~\ref{fig:barsalolnew30}) shows dynamics dominated by
references to years. (The first version is similar. 
For analogous figures, see the supplementary S1--S4 Files.) 
Eight of the top ten
contributions to the divergence between those decades are due to
increased relative frequencies of use of each of years between 1940
and 1949, their contribution decreasing chronologically,
and the other two top ten words are the last two years of the
previous decade (``1948'' and ``1949'' appear at ranks 15 and 34, respectively).
The last three years in the 1920s also appear by way of decreased
relative frequency of use in the top 60 contributions.
Other notable differences include:
\begin{itemize}
\item 
  The 11th
  highest contribution is from ``war,'' which increased in relative
  frequency. 
\item 
  ``Hitler'' and ``Nazi'' (increased relative frequencies)
  are ranked 18th and 26th, respectively. 
\item 
  Parentheses (13th and 14th)
  show increased relative frequencies of use. 
\item 
  Personal pronouns show
  decreased relative frequencies of use. 
\item 
  The word ``King'' (41st) also
  shows a decreased relative frequency, possibly due to the British line
  of succession.
\end{itemize}

\subsubsection{Second unfiltered English data set: 1950s versus 1980s}
\label{subsubsec:wordshifts-unfeng2-1950s-1980s}

\begin{itemize}
\item 
  The top two contributions between the 1950s and the 1980s (see
  Fig.~\ref{fig:barsallnew50}) in the English data set are both parentheses,
  which show dramatically increased relative frequencies of
  use. 
\item 
  Combined with increased relative frequencies for the colon (4th),
  solidus/virgule (or forward slash) (14th), ``computer'' (32nd), 
  and square brackets (58th
  and 59th), this suggests that the primary changes between the 1950s
  and the 1980s are due specifically to computational sources. 
\item 
  Other
  technical words showing noticeable increases include ``model'' (34th),
  ``data'' (35th), ``percent'' and the percentage sign (37th and 39th),
  ``Figure'' (40th), ``technology'' (51st), and ``information''
  (56th). 
\item 
  Similarly to the divergence between the 1930s and 1940s, 19
  out of the top 30 places are accounted for by increased relative
  frequencies of use in years between 1968 and 1980. 
\item 
  The words ``the''
  (3rd), ``of'' (8th), and ``which'' (16th) all decrease noticeably in
  relative frequency and are the highest ranked alphabetical
  1-grams. 
\item 
  Unlike the divergence between the 1930s and 1940s, only
  masculine pronouns show decreases in the top 60, while ``women''
  (55th) increases.
\end{itemize}

\subsubsection{First English fiction data set: 1930s versus 1940s}
\label{subsubsec:wordshifts-engfic1-1930s-1940s}

The first version of English Fiction shows similar dynamics to the
second version of the unfiltered data set between the 1930s and the
1940s (see Fig.~\ref{fig:barsficold30}) with yearly mentions dominating
the ranks. Some exceptions include:
\begin{itemize}
\item 
  ``Lanny'' rising in rank from 49th
  to 8th.
\item 
  Parentheses falling from 13th and 14th to 36th and 37th.
  ``ml'' (increased relative frequency of use in the 1940s) falling from
  31st to 55th.
\item 
  ``radio'' (with increased relative frequency) rising
  from 51st to 30th. 
\item
  ``King'' is no longer in the top 60
  contributions. 
\item
  ``patient'' enters the top 60 (ranked 51st).
\end{itemize}

\subsubsection{First English fiction data set: 1950s versus 1980s}
\label{subsubsec:wordshifts-engfic1-1950s-1980s}

This similarity between the original English Fiction data set and the
unfiltered data set also appears in the divergence between the 1950s
and the 1980s (see Fig.~\ref{fig:barsficold50}) with parentheses and years
dominating. Moreover, ``patients'' ranks 13th (with increased relative
frequency of use) despite not appearing in the top 60 for the
unfiltered data set. These observations, combined with increases in
``levels'' (47th), ``drug'' (51st), ``response'' (55th), and
``therapy'' (56th) demonstrate the original fiction data set is
strongly influenced by medical journals. Therefore, this data set
cannot be considered as primarily fiction despite the label.

\begin{figure}[tbp!]
  \begin{center}
    \includegraphics[width=0.48\textwidth]{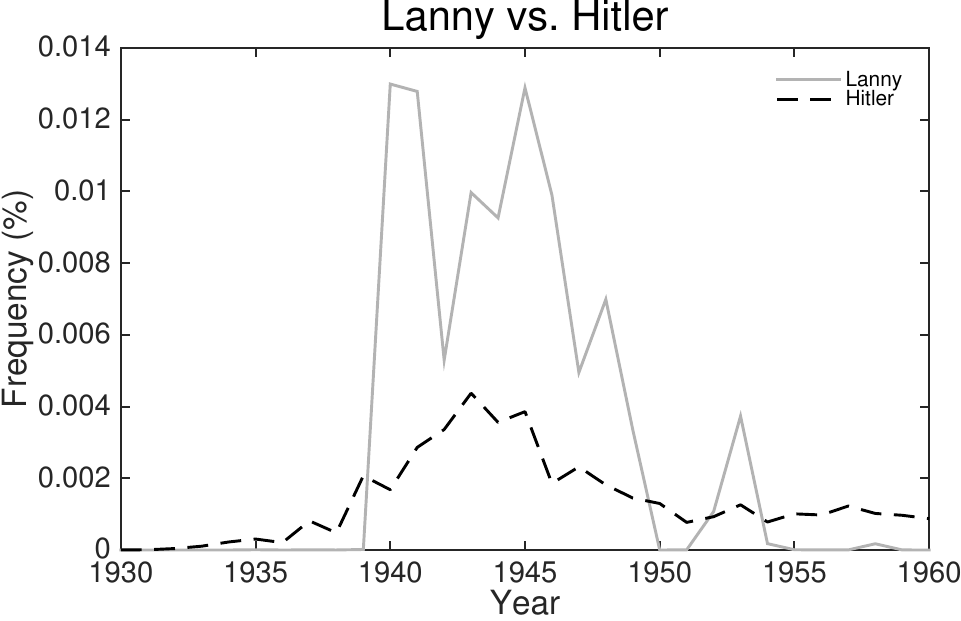}
  \end{center}
  \caption{ 
    Upton Sinclair wrote 11 Lanny Budd novels set during World
    War II. The first of these was published in 1940, and the last was
    published in 1953. The net effect of Sinclair's efforts is that his
    character appears much more frequently in the English Fiction
    (Version 2) data set than Hitler during most of the war. This
    demonstrates the potential impact of a single prolific author on the
    corpus.
  }
  \label{fig:lannyhitlerfig}
\end{figure}

\subsubsection{Second English fiction data set: 1930s versus 1940s}
\label{subsubsec:wordshifts-engfic2-1930s-1940s}

Fortunately, the same is not true for the second version of the
English Fiction data set. This is quickly apparent upon inspection of
the two greatest contributions to the divergence between the 1930s and
the 1940s (see Fig.~\ref{fig:barsficnew30}). 
The first of these is due to
a dramatic increase in the relative frequencies of use of quotation
marks, which implies increased dialogue. The second is the
name ``Lanny'' in reference to the recurring character Lanny Budd from
11 Upton Sinclair novels published between 1940 and 1953. 
``Budd''
ranks 11th in the chart ahead of ``Hitler'' (13th). 
The normalized frequency
series for ``Lanny'' and ``Hitler'' provided in
Fig.~\ref{fig:lannyhitlerfig} demonstrate that Lanny received
more mention than Hitler during this time period. 
The chart
is littered with the names of fictional characters:
\begin{itemize}
\item 
Studs Lonigan, the
1930s protagonist of a James T. Farrel trilogy, secures the 12th
spot. (Naturally, he is mentioned fewer times during the 1940s.) 
\item 
Dinny
Cherrel from the 1930s \textit{The Forsyte Saga} by John Galsworthy
secures rank 15. 
\item 
Wang Yuan from the 1930s \textit{The House of Earth}
trilogy by Pearl S. Buck ranks 17th and 37th. 
\item 
Detective Bill Weigand,
a recurring character created by Richard Lockridge in the 1940s,
secures rank 33.
\item 
The eponymous, original Asimov robot from the 1940 short story,
``Robbie,'' ranks 19th. 
\item 
``Mama'' (ranked 48th) is none other than the
subject of \textit{Mama's Bank Account}, published in 1943 by Kathryn
Forbes. 
\item 
``Saburov'' (ranked 22nd) from \textit{Days and Nights} by
Konstantin Simonov and ``Diederich'' (ranked 45th) from \textit{Der
  Untertan} by Heinrich Mann are subjects of works translated into
English in the 1940s. 
\end{itemize}

We note that while Marcel Proust (56th and 33rd), who died
in 1922 may be present in the 1940s due to letters translated by Mina
Curtiss in 1949 or other references not technically
fiction.
Similarly, ``B.M.'' (18th) may refer to the author
B.~M.~Bower.
Thus, the vast majority of prominent words in the word shift may
be traced not only to authors of fiction, but to the content of their
work. Moreover, the greatest contributions to divergence appear to
correspond to the most prolific authors, particularly Upton Sinclair.

\begin{figure}[tbp!]
  \begin{center}
    \includegraphics[width=0.48\textwidth]{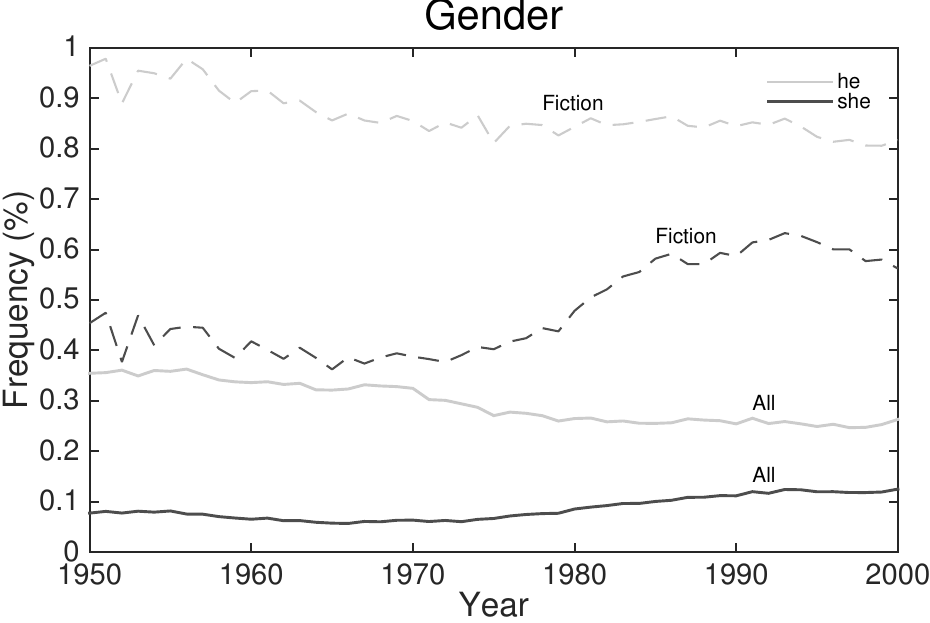}
  \end{center}
  \caption{
    Time series for ``he'' and ``she'' for Version 2. The
    unfiltered normalized frequencies are given by the solid curve. 
    Normalized frequencies in
    fiction are given by the dashed curve. These personal pronouns are
    more common in fiction. The pronoun ``she'' gains popularity through
    the 1990s in both data sets, with a more pronounced growth
    in fiction.
  }
  \label{fig:heshefig}
\end{figure}

\subsubsection{Second English fiction data set: 1930s versus 1940s}
\label{subsubsec:wordshifts-engfic2-1950s-1980s}

While there are no names of characters in the top divergences between
the 1950s and the 1980s, the updated fiction data set
(Fig.~\ref{fig:barsficnew50}) displays far more variety than the original
version, including:
\begin{itemize}
\item 
  Decreases in relative frequencies of masculine
  pronouns---e.g., ``he'' (rank 19) and ``himself'' (rank 48)---and
  corresponding increases for feminine pronouns---e.g., ``her'' (3rd),
  ``she'' (5th), and ``She'' (6th).  We present times series for ``he'' and
  ``she'' in Fig.~\ref{fig:heshefig}.
\item 
  An increase in relative frequencies of contractions (see ranks 9, 15, and
  21).
\item 
  A decrease in ``shall'' (16th) and ``must'' (49th), and a variety
  of increased profanity (particularly ranks 33 and 51).
\item
  Decreases in ``Mr.'' (10th) and ``Mrs.'' (17th).
\item 
  Various shifts in punctuation, particularly fewer
  semicolons (1st) and more periods (2VD). Quotation (11th) and question
  (18th) marks both see increased relative frequencies of use in the
  1980s, and the four-period ellipsis (20th) loses ground to the
  three-period version (22TD). 
\end{itemize}

\begin{figure}[tbp!]
  \begin{center}
    \includegraphics[width=0.48\textwidth]{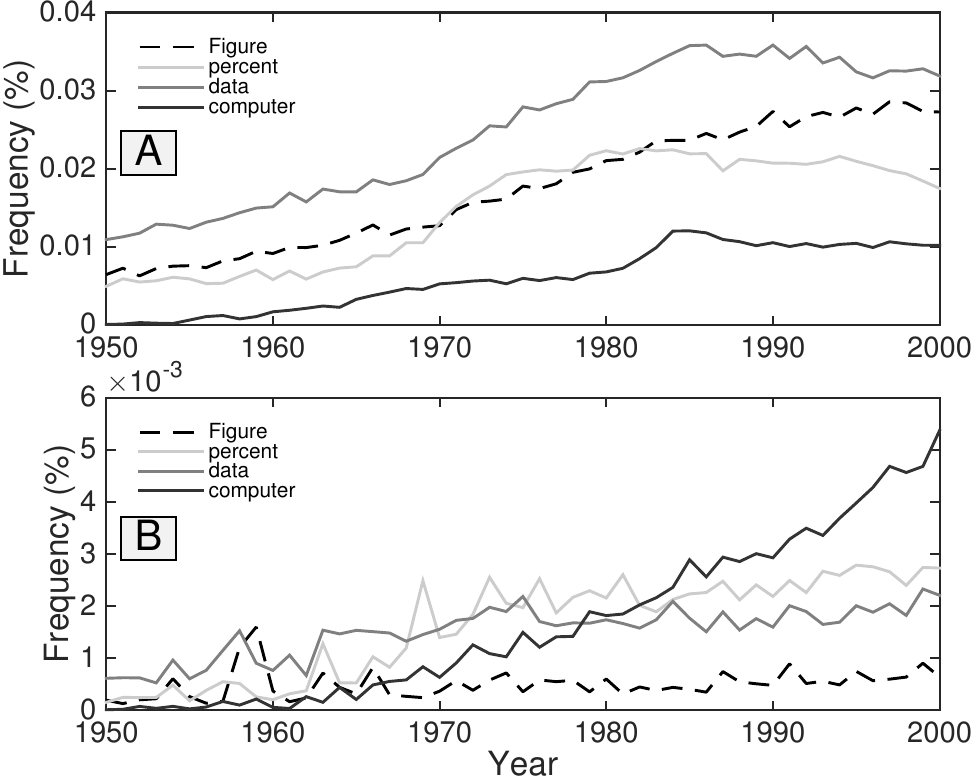}
  \end{center}
  \caption{ 
    Time series of technical terms from Version 2: (a) English
    all, (b) English fiction.
    In the unfiltered data set, these technical terms appear frequently
    and increase in usage though the 1980s. In fiction, technical terms
    show up far less frequently and remain relatively stable in usage
    with the notable exception of ``computer,'' which has been gradually
    gaining popularity since the 1960s. }
  \label{fig:moreseries}
\end{figure}

\subsection{The rise of  scientific literature in the Google Books corpus}
\label{subsec:science}

As our JSD analysis has shown above,
the unfiltered English data sets feature more general scientific
terms and we compare ``percent,'' ``data,'' ``Figure,'' and
``model'' in Fig.~\ref{fig:moreseries}. 
The original fiction data set also features
these, but also places ``patients,'' ``drug,'' ``response,'' and
``therapy'' among the top 60 contributions.  The primary
difference between the unfiltered and original fiction data sets in
the 1980s (compared to the 1950s) appears to consist of the nature of
journals sampled. The unfiltered components predicted and observed for
this particular data set seem to be dominated by medical journals.

\begin{figure}[tbp!]
  \begin{center}
    \includegraphics[width=0.48\textwidth]{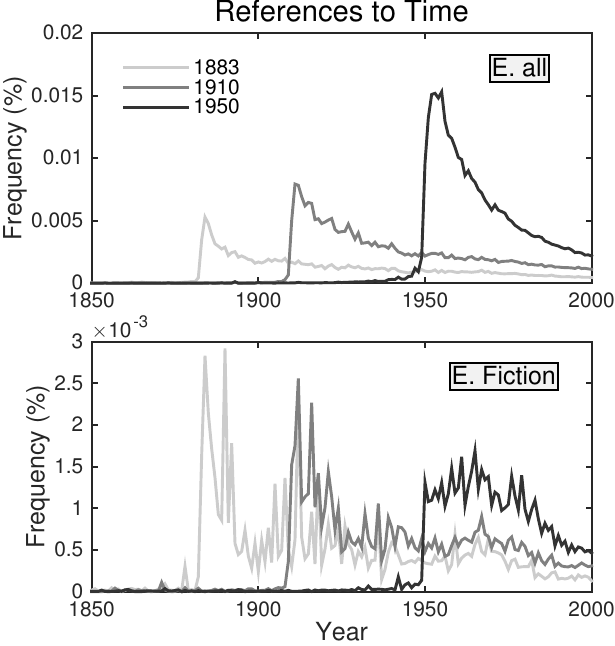}
  \end{center}
  \caption{
    Normalized frequencies of references to years. The top panel
    resembles a figure from~\cite{michel2011quantitative} using
    unfiltered data from English Version 2. (The cited paper uses
    Version 1.) Note the characteristic rapid rises and gradual
    declines, as well as the increasing peaks in yearly
    references. However, while the characteristic shape is still present
    in fiction (Version 2, bottom)---at much reduced levels---the peaks
    do not rise. The rising effect is likely due to citations from
    scientific texts.}
  \label{fig:yearsfig}
\end{figure}

As well as having more mentions of time and technical terms (and
parentheses) in the 1980s than in the 1950s, both unfiltered versions
and the first fiction data set include both ``et'' and ``al'' with
greater relative frequency in the 1980s. Perhaps more importantly,
years do not have a large effect on the dynamics in the second English
Fiction data set. We see in Fig.~\ref{fig:yearsfig} that while peaks for
years rise in the unfiltered data, they do not in fiction. The absence
of rising peaks in fiction strongly suggests the rise in peak relative
frequencies of years in the larger data set is due to a citation bias
in the unfiltered data set from high sampling of scientific
journals. This bias casts strong doubt on conclusions that we as a culture
forget things more quickly than we once did based on the observation
that half-lives for mentions of a given year decline over
time~\cite{michel2011quantitative}.

The exponential rise in scientific literature is not a new
phenomenon, and as de Solla Price stated in
1963~\cite{price1963a}~(p. 81) 
when discussing the half-lives for
citations of scientific literature, ``In fields embarrassed by an
inundation of literature there will be a tendency to bury as much of
the past as possible and to cite older papers less often than is their
statistical due.'' 
It would seem that an explanation
for declining half-lives in the mentions of years 
lies in the dynamics of the memory of scientific discoveries
rather than that of culture.

For the second fiction data set, we observe
in Fig.~\ref{fig:moreseries}B, 
that ``computer'' gains popularity in the fiction data set despite other technical words
remaining relatively steady in usage, as we might expect. 
This should be encouraging for anyone attempting to
analyze colloquial English, despite the prolificacy bias apparent from
the authors such as Upton Sinclair.

In the supplementary S1--S4 Files,
we include the top 60 contributions to divergences between each pair of
the 20 decades in each of the four data sets analyzed in this
paper. In total, 760 figures are included (190 per data set) for a
grand total of 45,600 contributions. We highlight some of these here.

\begin{itemize}
\item 
  For divergences to and from the first decade of the 1800s, many of
the contributions are due to a reduction of optical character
recognition confusion between the letters `f' and `s'. For example, in
the second unfiltered data set between the 1800s and 1810s, the top
two contributions are due to reductions in ``fame'' and ``os,''
respectively. The word ``same'' (ranked 11th) is the first increasing
contribution. Decreased relative frequencies of ``os,'' ``￼sirst,''
``thofe,'' ``fo,'' ``fay,'' ``cafe,'' ``fays,'' ``fome,'' and ``faid''
(ranks 3 through 10, respectively) and ``lise'' (12th) all suggest
digital misreadings of both 'f' and the long~`s'. (The 13th
contribution is ``Napoleon,'' who is mentioned with greater relative
frequently in the 1810s.)
\item Contributions between the 1830s and the 1860s in the second
unfiltered data set highlight the American Civil War and its
aftermath. ``State'' (11th), ``General'' (19th), ``States'' (20th),
``Union'' (37th), ``Confederate'' (48th), ``Government'' (52nd),
``Federal'' (56th), and ``Constitution'' (59th) all show increased
relative frequency of use. Religious terms tend to decline during this
period---e.g., ``church'' (14th), ``God'' (24th), and ``religion''
(58th).
\item Between the 1940s and 1960s, the second unfiltered dataset shows
increases for ``nuclear'' (43rd), ``Vietnam'' (47th), and
``Communist'' (50th). The relative frequency of ``war'' (25th)
decreases substantially. Meanwhile in fiction, ``Lanny'' (5th)
declines, while ``television'' (38th) and the Hardy Boys (``Hardy''
ranks 51st) appear with greater relative frequencies.
\item Between the 1960s and 1970s, the second fiction data set is
strongly affected by ``Garp'' (\textit{The World According to Garp} by
John Irving, 1978) at rank 19, increased relative frequencies of
profanity (ranks 27, 33, and 38), and increased mentions of ``Nixon''
(41st) and ``Spock'' (47th, likely due to ``Star Trek'' novels).
\item Between the 1980s and 1990s, the second fiction set shows
increased relative frequencies of use of the words ``gay'' (15th),
``lesbian'' (19th), ``AIDS'' (24th), and ``gender'' (27th). Female
pronouns (2nd, 8th, and 9th) show increased relative frequencies of
use in continuance of Fig.~\ref{fig:barsficnew50}.
\end{itemize}

\bigskip

\section{Concluding remarks} 
\label{sec:conc}

Based on our introductory remarks
and ensuing detailed analysis, it should now be 
clear that the contents of the Google Books corpus 
do not represent an unbiased sampling of publications.
Beyond being library-like,
the evolution of the corpus throughout the 1900s is increasingly 
dominated by scientific publications
rather than popular works.
We have shown that even the first data set
specifically labeled as fiction appears to be saturated with medical
literature.

When examining these data sets in the future, it will therefore be necessary
to first identify and distinguish the popular and scientific components in order to form a
picture of the corpus that is informative about cultural and
linguistic evolution. 
For instance, one should ask how much of any observed gender 
shift in language reflects word choice in popular works 
and how much is due to changes in scientific norms, as well as
which might precede the other if they are somewhat in balance.

Even if we are able to restrict our focus to popular works by
appropriately filtering scientific terms,
the library-like nature of the Google Books corpus 
will mean the resultant normalized frequencies of words cannot be a direct measure 
of the ``true'' cultural popularity of those words as they are read (again, Frodo). 
Secondarily, not only will there
be a delay between changes in the public popularity of words and 
their appearance in print,
normalized frequencies will also be affected by the prolificacy of the authors.
In the  case of Upton Sinclair's Lanny Budd, a fictional character was 
vaulted to the upper echelons of words affecting divergence
(even surpassing Hitler) by virtue of appearing as the protagonist in 11 novels between 1940 and 1953. 
Google Books is at best a limited proxy for social information after the fact.

The Google Books corpus's beguiling power to immediately quantify a vast range
of linguistic trends 
warrants a very cautious approach to any effort to extract scientifically meaningful results. 
Our analysis provides a possible framework for improvements to previous and future works
which, if performed on English data, ought to focus solely
on the second version of the English Fiction data set, 
or otherwise properly account for the biases of the unfiltered corpus.

\acknowledgments

\revtexonly{PSD was supported by NSF CAREER Award \# 0846668.}

\clearpage


\begin{thebibliography}{10}
\expandafter\ifx\csname bibnamefont\endcsname\relax
  \def\bibnamefont#1{#1}\fi
\expandafter\ifx\csname bibfnamefont\endcsname\relax
  \def\bibfnamefont#1{#1}\fi
\expandafter\ifx\csname url\endcsname\relax
  \def\url#1{\texttt{#1}}\fi
\expandafter\ifx\csname urlprefix\endcsname\relax\def\urlprefix{URL }\fi
\providecommand{\bibinfo}[2]{#2}
\providecommand{\eprint}[2][]{\url{#2}}

\bibitem{michel2011quantitative}
\bibinfo{author}{\bibfnamefont{J.-B.} \bibnamefont{Michel}},
  \bibinfo{author}{\bibfnamefont{Y.~K.} \bibnamefont{Shen}},
  \bibinfo{author}{\bibfnamefont{A.~P.} \bibnamefont{Aiden}},
  \bibinfo{author}{\bibfnamefont{A.}~\bibnamefont{Veres}},
  \bibinfo{author}{\bibfnamefont{M.~K.} \bibnamefont{Gray}},
  \bibinfo{author}{\bibfnamefont{J.~P.} \bibnamefont{Pickett}},
  \bibinfo{author}{\bibfnamefont{D.}~\bibnamefont{Hoiberg}},
  \bibinfo{author}{\bibfnamefont{D.}~\bibnamefont{Clancy}},
  \bibinfo{author}{\bibfnamefont{P.}~\bibnamefont{Norvig}},
  \bibinfo{author}{\bibfnamefont{J.}~\bibnamefont{Orwant}}, \emph{et~al.},
  \bibinfo{journal}{science}
  \textbf{\bibinfo{volume}{331}}(\bibinfo{number}{6014}), \bibinfo{pages}{176}
  (\bibinfo{year}{2011}).

\bibitem{lin2012syntactic}
\bibinfo{author}{\bibfnamefont{Y.}~\bibnamefont{Lin}},
  \bibinfo{author}{\bibfnamefont{J.-B.} \bibnamefont{Michel}},
  \bibinfo{author}{\bibfnamefont{E.~L.} \bibnamefont{Aiden}},
  \bibinfo{author}{\bibfnamefont{J.}~\bibnamefont{Orwant}},
  \bibinfo{author}{\bibfnamefont{W.}~\bibnamefont{Brockman}}, \bibnamefont{and}
  \bibinfo{author}{\bibfnamefont{S.}~\bibnamefont{Petrov}}, in
  \emph{\bibinfo{booktitle}{Proceedings of the ACL 2012 System Demonstrations}}
  (\bibinfo{organization}{Association for Computational Linguistics},
  \bibinfo{year}{2012}), pp. \bibinfo{pages}{169--174}.

\bibitem{salganik2006a}
\bibinfo{author}{\bibfnamefont{M.~J.} \bibnamefont{Salganik}},
  \bibinfo{author}{\bibfnamefont{P.~S.} \bibnamefont{Dodds}}, \bibnamefont{and}
  \bibinfo{author}{\bibfnamefont{D.~J.} \bibnamefont{Watts}},
  \bibinfo{journal}{Science} \textbf{\bibinfo{volume}{311}},
  \bibinfo{pages}{854} (\bibinfo{year}{2006}).

\bibitem{frodo-google-ngrams2015a}
\emph{\bibinfo{title}{Google {N}gram {V}iewer: `{F}rodo', 1800--2000 in
  {E}nglish {F}iction.}},
  \bibinfo{note}{\texttt{https://books.google.com/ngrams/graph?}
  \texttt{content=Frodo\&year\_start=1800}
  \texttt{\&year\_end=2000\&corpus=16\&smoothing=1}; Accessed April 25, 2015}.

\bibitem{twenge2012increases}
\bibinfo{author}{\bibfnamefont{J.~M.} \bibnamefont{Twenge}},
  \bibinfo{author}{\bibfnamefont{W.~K.} \bibnamefont{Campbell}},
  \bibnamefont{and} \bibinfo{author}{\bibfnamefont{B.}~\bibnamefont{Gentile}},
  \bibinfo{journal}{PloS one} \textbf{\bibinfo{volume}{7}}(\bibinfo{number}{7})
  (\bibinfo{year}{2012}).

\bibitem{twenge2012male}
\bibinfo{author}{\bibfnamefont{J.~M.} \bibnamefont{Twenge}},
  \bibinfo{author}{\bibfnamefont{W.~K.} \bibnamefont{Campbell}},
  \bibnamefont{and} \bibinfo{author}{\bibfnamefont{B.}~\bibnamefont{Gentile}},
  \bibinfo{journal}{Sex roles}
  \textbf{\bibinfo{volume}{67}}(\bibinfo{number}{9-10}), \bibinfo{pages}{488}
  (\bibinfo{year}{2012}).

\bibitem{greenfield2013changing}
\bibinfo{author}{\bibfnamefont{P.~M.} \bibnamefont{Greenfield}},
  \bibinfo{journal}{Psychological science}
  \textbf{\bibinfo{volume}{24}}(\bibinfo{number}{9}), \bibinfo{pages}{1722}
  (\bibinfo{year}{2013}).

\bibitem{petersen2012statistical}
\bibinfo{author}{\bibfnamefont{A.~M.} \bibnamefont{Petersen}},
  \bibinfo{author}{\bibfnamefont{J.}~\bibnamefont{Tenenbaum}},
  \bibinfo{author}{\bibfnamefont{S.}~\bibnamefont{Havlin}}, \bibnamefont{and}
  \bibinfo{author}{\bibfnamefont{H.~E.} \bibnamefont{Stanley}},
  \bibinfo{journal}{Scientific reports} \textbf{\bibinfo{volume}{2}}
  (\bibinfo{year}{2012}).

\bibitem{gerlach2013stochastic}
\bibinfo{author}{\bibfnamefont{M.}~\bibnamefont{Gerlach}} \bibnamefont{and}
  \bibinfo{author}{\bibfnamefont{E.~G.} \bibnamefont{Altmann}},
  \bibinfo{journal}{Physical Review X}
  \textbf{\bibinfo{volume}{3}}(\bibinfo{number}{2}), \bibinfo{pages}{021006}
  (\bibinfo{year}{2013}).

\bibitem{petersen2012languages}
\bibinfo{author}{\bibfnamefont{A.~M.} \bibnamefont{Petersen}},
  \bibinfo{author}{\bibfnamefont{J.~N.} \bibnamefont{Tenenbaum}},
  \bibinfo{author}{\bibfnamefont{S.}~\bibnamefont{Havlin}},
  \bibinfo{author}{\bibfnamefont{H.~E.} \bibnamefont{Stanley}},
  \bibnamefont{and} \bibinfo{author}{\bibfnamefont{M.}~\bibnamefont{Perc}},
  \bibinfo{journal}{Scientific reports} \textbf{\bibinfo{volume}{2}}
  (\bibinfo{year}{2012}).

\bibitem{bentley2014books}
\bibinfo{author}{\bibfnamefont{R.~A.} \bibnamefont{Bentley}},
  \bibinfo{author}{\bibfnamefont{A.}~\bibnamefont{Acerbi}},
  \bibinfo{author}{\bibfnamefont{P.}~\bibnamefont{Ormerod}}, \bibnamefont{and}
  \bibinfo{author}{\bibfnamefont{V.}~\bibnamefont{Lampos}},
  \bibinfo{journal}{PloS ONE}
  \textbf{\bibinfo{volume}{9}}(\bibinfo{number}{1}), \bibinfo{pages}{e83147}
  (\bibinfo{year}{2014}).

\bibitem{koplenig2015a}
\bibinfo{author}{\bibfnamefont{A.}~\bibnamefont{Koplenig}},
  \bibinfo{journal}{Digital Scholarship in the Humanities}
  (\bibinfo{year}{2015}), \bibinfo{note}{in press}.

\bibitem{kullback1951information}
\bibinfo{author}{\bibfnamefont{S.}~\bibnamefont{Kullback}} \bibnamefont{and}
  \bibinfo{author}{\bibfnamefont{R.~A.} \bibnamefont{Leibler}},
  \bibinfo{journal}{The Annals of Mathematical Statistics} pp.
  \bibinfo{pages}{79--86} (\bibinfo{year}{1951}).

\bibitem{shannon2001mathematical}
\bibinfo{author}{\bibfnamefont{C.~E.} \bibnamefont{Shannon}},
  \bibinfo{journal}{ACM SIGMOBILE Mobile Computing and Communications Review}
  \textbf{\bibinfo{volume}{5}}(\bibinfo{number}{1}), \bibinfo{pages}{3}
  (\bibinfo{year}{2001}).

\bibitem{lin1991divergence}
\bibinfo{author}{\bibfnamefont{J.}~\bibnamefont{Lin}},
  \bibinfo{journal}{Information Theory, IEEE Transactions on}
  \textbf{\bibinfo{volume}{37}}(\bibinfo{number}{1}), \bibinfo{pages}{145}
  (\bibinfo{year}{1991}).

\bibitem{dodds2011e}
\bibinfo{author}{\bibfnamefont{P.~S.} \bibnamefont{Dodds}},
  \bibinfo{author}{\bibfnamefont{K.~D.} \bibnamefont{Harris}},
  \bibinfo{author}{\bibfnamefont{I.~M.} \bibnamefont{Kloumann}},
  \bibinfo{author}{\bibfnamefont{C.~A.} \bibnamefont{Bliss}}, \bibnamefont{and}
  \bibinfo{author}{\bibfnamefont{C.~M.} \bibnamefont{Danforth}},
  \bibinfo{journal}{PLoS ONE} \textbf{\bibinfo{volume}{6}},
  \bibinfo{pages}{e26752} (\bibinfo{year}{2011}), \bibinfo{note}{arXiv version
  available at
  \href{http://arxiv.org/abs/1101.5120v4}{http://arxiv.org/abs/1101.5120v4}}.

\bibitem{price1963a}
\bibinfo{author}{\bibfnamefont{D.~J.} \bibnamefont{de~Solla~Price}},
  \emph{\bibinfo{title}{Little Science, Big Science}}
  (\bibinfo{publisher}{Columbia University Press}, \bibinfo{address}{New York},
  \bibinfo{year}{1963}).

\end{thebibliography}
\end{document}